\documentclass[twoside]{article}
\usepackage[utf8]{inputenc}
\usepackage[english]{babel}
\usepackage{moreverb,url}
\usepackage{amsmath}
\usepackage{amssymb}
\usepackage{booktabs} %
\usepackage{graphicx}
\usepackage[all]{nowidow}
\usepackage[utf8]{inputenc}
\usepackage{multicol}

\usepackage[colorlinks,bookmarksopen,bookmarksnumbered,citecolor=red,urlcolor=red]{hyperref}
\usepackage{RR}
\usepackage{cleveref}

\newcommand\BibTeX{{\rmfamily B\kern-.05em \textsc{i\kern-.025em b}\kern-.08em
T\kern-.1667em\lower.7ex\hbox{E}\kern-.125emX}}

\usepackage{xcolor}
\usepackage{xspace}
\usepackage[autolanguage]{numprint}

\RRNo{9377}
\RRdate{November 2020}

 \definecolor{monrouge}{HTML}{AD4B42}
 \definecolor{monbleu}{HTML}{4F6A9C}
 \definecolor{monvert}{HTML}{82A35D}
 \definecolor{monjaune}{HTML}{DCB355}
 \definecolor{cyan}{HTML}{00AAAA}
 \newcommand{\sebastian}[1]{{\color{monrouge}{\bf SEBASTIAN:} \small #1}}

\newcommand{\blindcomment}[1]{}
 \newcommand{\hide}[1]{}

\newcommand{\melissa}{Melissa-DA\xspace}
\newcommand{\meli}{melissa}

\definecolor{blue}{HTML}{1F77B4}
\definecolor{orange}{HTML}{FF7F0E}
\definecolor{green}{HTML}{2CA02C}

\newcommand{\uu}[2]{{\numprint{#1}}\,{$\text{{#2}}$}}

\RRtitle{An elastic framework for ensemble-based large-scale data~assimilation}
\RRetitle{An elastic framework for ensemble-based large-scale data~assimilation}
\RRauthor{Sebastian Friedemann \and Bruno Raffin}

\RRabstract{

Prediction of chaotic systems relies on a floating fusion of sensor data (observations) with a numerical model to decide on a good system trajectory and to compensate nonlinear feedback effects. Ensemble-based data assimilation~(DA) is a major method for this concern depending on propagating an ensemble of perturbed model realizations.

In this paper we develop an elastic, online, fault-tolerant and modular framework called \melissa for large-scale ensemble-based DA.
\melissa allows elastic addition or removal of compute resources for state propagation at runtime.
Dynamic load balancing based on list~scheduling ensures efficient execution. Online processing of the data  produced by  ensemble members enables to avoid the I/O~bottleneck of  file-based approaches.
Our implementation embeds the PDAF parallel DA engine, enabling the use of various DA methods.  \melissa can support  extra ensemble-based DA methods by implementing  the transformation of member background states into analysis states.
Experiments confirm the excellent scalability of \melissa, running on up to \numprint{16240} cores, to propagate \numprint{16384} members for a regional hydrological critical zone assimilation relying on the ParFlow model on a domain with about \uu{4}{M} grid cells.

}

\RRkeyword{Data~Assimilation, Ensemble~Kalman~Filter, Ensemble, Multi~Run~Simulations, Elastic, Fault~Tolerant, Online, In~Transit Processing, Master/Worker}

\RRprojet{DataMove}
\URRhoneAlpes
\RCGrenoble

\begin{document}

\makeRR
\section{Introduction}\label{sec:intro}
Numerical models of highly nonlinear (chaotic) systems (e.g., the atmospheric, the oceanic or the ground water flow) are extremely sensitive to input variations. The goal of data assimilation (DA)  is to  reduce the result uncertainty by correcting the model trajectory  using  observation data.

Two  main  approaches are used for  DA,  variational and  statistical
~\cite{Asch-DABook-2016,lorenz_deterministic_1963}. In  this paper  we focus  on statistical ensemble-based DA, where an ensemble of several  model instances is executed to estimate the model error against the observation error. Various methods exist for that purpose like the classical Ensemble Kalman Filter (EnKF) we use for  experiments in this paper.

Combining large numerical models and ensemble-based DA requires execution on supercomputers. Recent models have millions of degrees of freedom, but only hundreds of ensemble members are used for operational DA. Thus today's approaches suffer from undersampling: to minimize and to estimate the sampling error, much larger ensembles with thousands of members would be necessary. Current approaches rely either on files to aggregate the ensemble results and send back corrected states or  on the  MPI message passing programming paradigm  to  harness the members and the assimilation process in a  large monolithic code, but these approaches are not well-suited to  running ultra-large ensembles on the coming exascale computers.  Efficient execution at exascale requires mechanisms to allow dynamic adaptation to the context, including load balancing for limiting  idle time, elasticity to adapt to the machine availability, fault tolerance to recover from failures caused by numerical, software or hardware issues and direct communications between components instead of files to bypass the I/O performance bottleneck.

This paper proposes a novel architecture characterized  by its flexibility and modularity called \melissa. Experiments using  the ParFlow parallel hydrology  solver on a domain with about \uu{4}{M} grid cells as numerical model were run with up to \numprint{16384} members on \numprint{16240} cores. Each member itself was parallelized on up to \numprint{48} MPI ranks.  \melissa adopts a three-tier architecture based on a launcher in charge
of orchestrating the full application execution,  independent parallel runners in charge of
executing members up to the next assimilation update phase, and a parallel server that gathers and
updates member states to assimilate observation data into them.  The benefits of this framework include:
\begin{itemize}
   \item   \textbf{Elasticity:}  \melissa enables the dynamic adaptation of compute resource usage  according
  to  availability.   Runners  are independent and connect dynamically to the
parallel server when they start. They are submitted as independent jobs to the batch
scheduler. Thus, the number of concurrently running  runners  can vary during the course of  a study
to adapt to the availability of compute resources.

\item \textbf{Fault tolerance:}  \melissa's  asynchronous master/worker architecture supports a
  simple yet robust fault tolerance mechanism. Only some lightweight book-keeping and a few
  heartbeats as well as  checkpointing on the server side are required to detect issues and restart the server or the runners.

\item \textbf{Load balancing:}  The distribution of member states to runners is controlled by the server
  and defined dynamically according to a list~scheduling algorithm,  enabling to adjust the load of each runner according to the time required to propagate each member.

\item \textbf{Online processing:} Exchange of state variables between the different parts of the \melissa application and the different parts of an assimilation cycle happens fully online avoiding file system access and its latency.

\item \textbf{Communication/computation overlap:}  Communications between the server and the
  runners occur asynchronously, in parallel with computation, enabling an effective overlapping
  of computation and communication, improving the overall execution efficiency.

  \item   \textbf{Code modularity:}   \melissa enforces code modularity by design, leading to a clear separation of concern between models and  DA.  The runners execute the members, i.e., model instances, while the server, a separate code running in a different job, is concerned with  observation processing  and running the core DA algorithm.
\end{itemize}

After  a quick reminder about statistical data assimilation and especially the Ensemble Kalman Filter  (\autoref{sec:DA}), related work is
presented (\autoref{sec:relatedwork}). The proposed architecture and the \melissa framework are detailed  (\autoref{sec:melissa})  before  analyzing experimental results
(\autoref{sec:exp}). A conclusion closes the paper (\autoref{sec:conclusion}).

\section{Statistical data assimilation and the ensemble Kalman filter}\label{sec:DA}

\begin{figure}
  \centering
  \includegraphics[width=1\linewidth]{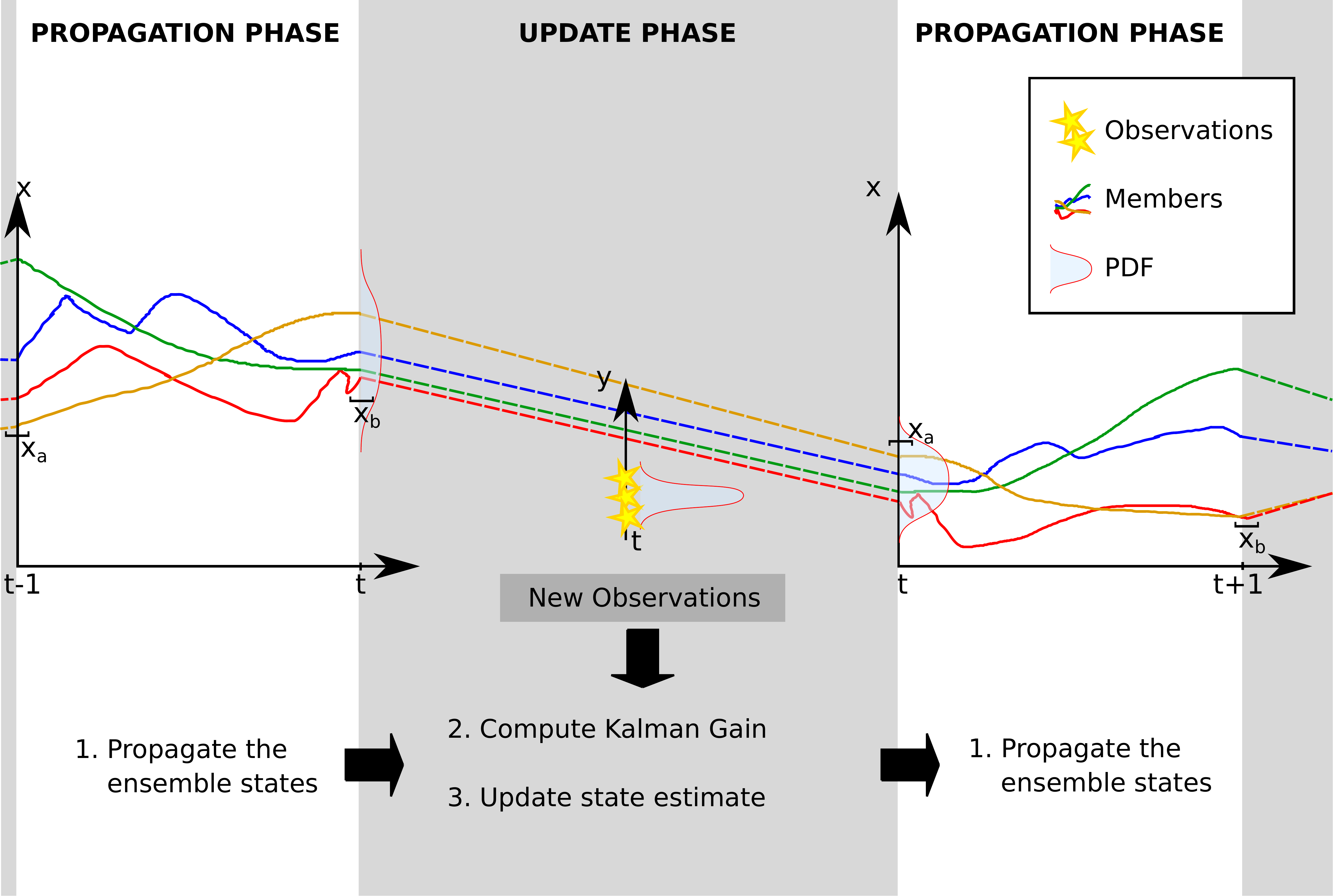}
\caption{The ensemble Kalman filter workflow.}
\label{fig:enkf}
\end{figure}

We survey the base concepts of ensemble-based data assimilation and the EnKF filter. Refer to~\cite{Asch-DABook-2016} and~\cite{Evensen-DABook-2009} for more details.

The goal of DA is to estimate the real state of the system as accurately as possible merging  model intermediate results and observations. This is critical in particular to models showing  chaotic or semichaotic behavior, where  small changes in the model input lead to large changes in the model output. The state estimate is called \textit{analysis~state}~$x_a$. It might be an estimate of, e.g., the present atmospheric state in Numerical Weather Prediction~(NWP) or the soil moisture field when calibrating hydrological ground water models using DA.  The analysis state $x_a$ is retrieved running a model (e.g., the atmospheric or hydrological model itself) and improving the model output using the observation vector $y$. Observation vectors in the geoscientific domain typically contain a mix of values from remote sensing instruments (satellites) and in~situ observations by ground based observatories, buoys, aircrafts etc.

Models operate on the system state~$x \in X \subseteq \mathbb{R}^N$ that cannot be directly observed. In the standard DA formalism, the model operator $\mathcal{M}$ fulfills the Markov property, taking the present state~$x_{t}$ as its \textit{only} input to produce the next state~$x_{t+1}$:
\begin{equation}
   x_{t+1} =  \mathcal{M}(x_t).
\end{equation}

To compare and weight an observation vector~$y \in Y \subseteq \mathbb{R}^K $, and system state~$x$, we rely on an observation operator~$\mathcal{H}$ that maps from model space to the observation space:
\begin{equation}
    \tilde{y} = \mathcal{H}(x).
\end{equation}

In the case where both, the model output probability density function (PDF) and the observation's PDF are known,  multiplying the two and normalizing in Bayesian manner provides the  analysis state PDF.
Ensemble-based statistical DA estimates the model PDF by sampling an ensemble of model states and propagating these
states through $\mathcal{M}$.

The \textit{Ensemble Kalman Filter} (EnKF) is the most studied statistical ensemble-based DA method. This is the filter  we rely on to validate the \melissa approach, and its  workflow  is common to many other DA filters.    The EnKF inherits the Kalman Filter~\cite{kalman_filter_standard-1960}, extending it to non-linear models~\cite{Asch-DABook-2016}. The different steps of EnKF assimilation cycles are the following (\autoref{fig:enkf}):
\begin{enumerate}
    \item An ensemble $X_a$ of $M$ states $(x_a^i | i \in [M])$, statistically representing the assimilated state, is propagated by the model $\mathcal{M}$.  The  obtained states  are  the  background states $(x_b^i | i \in [M])$. For the initial assimilation cycle $X_a$ is an ensemble of perturbed states. Later it is obtained from the previous assimilation cycle. \label{itm:firstenkf}
    \item The Kalman gain $K$ is calculated from the  ensemble covariance and the observation error $R$.
    \item Multiply the Kalman gain $K$ with the innovation $\Big(y-\mathcal{H}(x_b^i)\Big)$ and add to the background states: $x_a^i = x_b^i + K \cdot \Big(y-\mathcal{H}(x_b^i)\Big)$ to obtain the new ensemble analysis states $(x_a^i | i \in [M])$.
    \item Start over with the next assimilation cycle (step \ref{itm:firstenkf}).
\end{enumerate}

\section{Related work}\label{sec:relatedwork}

DA techniques  fall into two  main categories, variational  and statistical.  Variational  DA (e.g.,
3D-Var and 4D-Var)  relies on the minimization of  a cost function evaluating the  difference between the
model state and  the observations.  Minimizing the  cost function is done via   gradient descent that
requires  adjoints for  both the  model  and the  observation  operator.  This  approach is  compute
efficient, but the  adjoints are not always  available or may require significant  efforts not always
accessible.  Nowadays large-scale DA applications as, e.g., used by Numerical Weather Prediction (NWP)
operators typically rely on a variational DA.  For instance, the China Meteorological Administration
uses 4D-Var  to assimilate about \numprint{2.1} million daily observations  into a global weather  model with
more  than \numprint{7.7}  million  grid cells.  As  in 2019,  the DA  itself  is parallelized  on  up to  \numprint{1024}
processes~\cite{zhang_operational_2019}.

Statistical DA  takes a different approach  relying on an ensemble  run of the model  to compute an
estimator of some  statistical operators (co-variance matrix for EnKF,  PDF for particle filters).
This approach consumes more compute power
as the number  of members needs to be large enough for the
estimator  to be  relevant, but stands for its simplicity as  it only  requires the  model operator  without adjoint as for  the
variational approach.  But scaling the ensemble size can be challenging especially  when the model
is already time consuming and requires its own internal parallelization.

Two main approaches are used to handle.  Either members run independently producing files that are read for the update
phase, that itself  produces new files read by members for the next assimilation cycle.  This approach makes
for a flexible workflow  amenable to  fault tolerance and  adjustable resource  allocations. Using files for
data exchange simplifies the issue of data redistribution  especially in the case where the
parallelization level of a given member differs from the one implemented for the update (usually
called a N$\times$M data redistribution).  This approach is adopted by the EnTK  framework~\cite{Balasu-EnTK-IPDPS2018},
used to manage up to \numprint{4096} members for DA on a molecular dynamics application~\cite{balasubramanian-EnTk-2020} and by Toye et al.\ \cite{Toye-FTschdDA-JCS2018} assimilating oceanic conditions in the Red sea  with O(\numprint{1000}) members.
The OpenDA framework also supports this  {\it blackbox} model, relying on NetCDF files for data
exchange for  NEMO in~\cite{van_velzen_openda-nemo_2016}.  However relying on the machine I/O capabilities using files is a growing  performance bottleneck.  In 10 years the compute power made a leap by a $134\times$ factor, from \uu{1.5}{PFlop/s} peak on Roadrunner (TOP500 \#1 in 2008) to \uu{201}{PFlop/s} peak on Summit
($\#1$ in 2018), the I/O throughput for the same machines only increased by a $12\times$ factor, from \uu{204}{GB/s} to \uu{2500}{GB/s}. This trend is expected to continue at exascale. For instance the announced Frontier machine (2021) expected to reach \uu{1}{ExaFlop/s} should offer 5 to 10 times the compute power of Summit but only 2 to 4 times its I/O throughput\footnote{see \url{https://www.olcf.ornl.gov/frontier/}, retrieved the 03.11.2020}.

The other approach builds a large MPI application that embeds and  coordinates the propagation of the different members as
well as the update phase.  Data exchange relies on efficient MPI collective communications. The
drawback is the monolithic aspect of this very large application. The full  amount of
resources needed  must be allocated upfront and for the full duration of the execution. Making such
application fault-tolerant and  load balanced is challenging.   Ensemble runs are particularly
sensible to numerical errors as the ensemble size increases the probability that the model
may execute in numerical domains it has not been well tested with.
The frameworks DART \cite{anderson_data_2009}
and PDAF \cite{Nerger-PDAF-2013} are based on this approach.  PDAF for instance has been used for the ground water flow model TerrSysMP
using  EnKF with up to 256 members~\cite{kurtz_terrsysmppdaf_2016}. Other DA work rely on custom MPI codes. Miyoshi et al.\ \cite{miyoshi_10240-member_2014} ran  \numprint{10240} members with the LENKF filter on the K-computer in 2014.
In 2018, Berndt et al.\  assimilated
up to \numprint{4096} members with \numprint{262144} processors and a particle filter, and production use case with \numprint{1024} mebebrs  for wind power prediction over Europe~\cite{berndt_predictability_2018}.
Variational and statistical DA can also be  combined.  NWP actors like the ECMWF, the British Met office and the Canadian Meteorological Centre rely on such approaches using hundreds of ensemble members to improve predictions \cite{kooij-connally_technical_nodate,clayton_operational_2013,houtekamer_using_2019}.

Looking beyond  DA frameworks,  various  Python based frameworks are supporting the automatic
distribution of tasks, enabling to manage ensemble runs. Dask~\cite{matthew_rocklin-proc-scipy-2015}
is for instance used for  hyperparameter search with Scikit-Learn,   Ray~\cite{moritz_ray_2018} for
reinforcement learning~\cite{Liang-RLlib-2018}. But they do not support tasks that are built from legacy MPI
parallel codes. Other frameworks such as Radical-Pilot (the base framework of EnTK)~\cite{Paraskevakos-EnTK-ICPP2018}
or Parsl~\cite{babuji-parsl-2019} enable such features but are using files for data exchange. Other
domain specific frameworks like Dakota~\cite{Elwasif-Dakota-CSE2015},
Melissa~\cite{terraz-melissa-SC17} or Copernicus~\cite{Pronk-copernicus-gmx-SC11} enable more
direct support of parallel simulation codes but for patterns that require less  synchronizations than the ones
required for DA.

\section{\melissa architecture}\label{sec:melissa}

\subsection{Overview}

\begin{figure}[!ht]
  \centering
  \vspace{.5cm}
   \includegraphics[width=.8\linewidth]{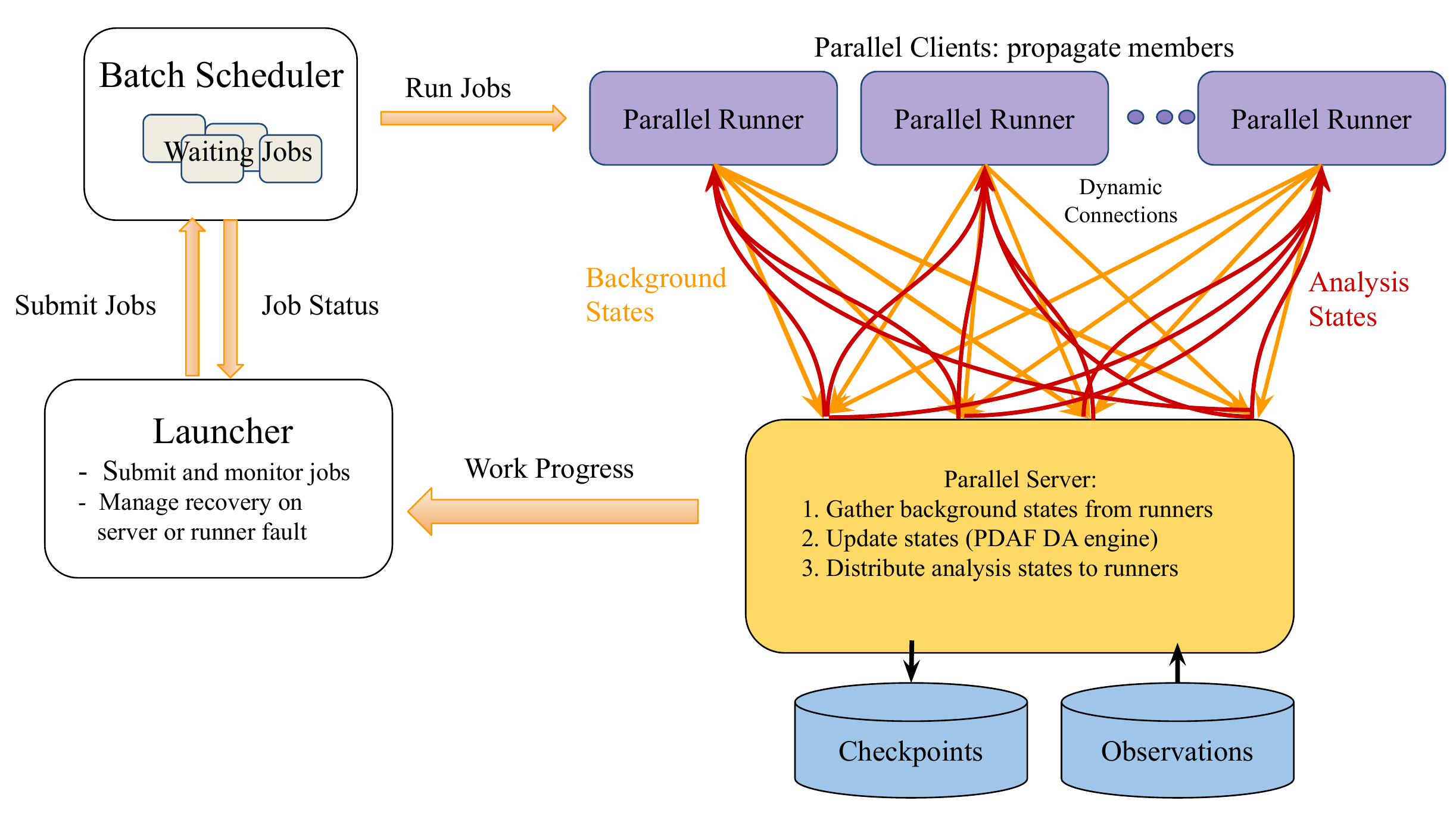}
\caption{ \melissa three-tier architecture. The launcher supervises the
execution in tight link with the batch scheduler. The job scheduler regulates the number of simulation jobs (runners) to run according to the machine availability,  leading to an elastic resource
usage.  The server  distributes the  members to propagate to
the connected runners dynamically for balancing their workload.
A fault tolerance mechanism automatically restarts failing runners or a failing server.}
\label{fig:launcher}
\end{figure}

\melissa  relies on an elastic and  fault-tolerant parallel
client/server communication scheme, inheriting a  three-tier architecture from Melissa~\cite{terraz-melissa-SC17}, (\autoref{fig:launcher}).
The  {\it server} gathers  background states from all ensemble members. New observations are assimilated into these background states to produce a new  analysis state for each member. These analysis states are distributed by the server   to the {\it runners} that take care of progressing the ensemble members up to the next assimilation cycle.
Member to runner distribution is adapted dynamically by the server according to the runner work loads,
following a list~scheduling algorithm.  The runners and the server are parallel codes that can run with different numbers of processes. They  exchange member states through  N~$\times$~M communication patterns for efficiency purpose.
These patterns map different runner domain parts stored by different runner ranks to server ranks. The server and the runners are launched independently in separate jobs by the machine batch scheduler. Runners connect dynamically to the server when starting.  This master/worker architecture enables an elastic resource usage, as the number of runners can dynamically evolve depending on the availability of compute resources.   The third tier is called  {\it   launcher}. The  launcher orchestrates  the execution, interacting with the supercomputer batch scheduler to request  resources to start new  jobs, kill jobs, monitor job statuses, and trigger job restarts  in case of failure.

\subsection{Member states} \label{sec:asim_only_part}

Often the observation operator $\mathcal{H}$ in a DA framework only applies to one part of
the whole model state vector. Thus the assimilated state vector denoted $x$ only needs to contain this part of the full model state. Other model variables that affect state propagation are left unchanged by the update phase. To account for this distinction   we split a model  state vector into two main parts, the {\it static state} and  the {\it dynamic state}, and we distinguish a sub-part of the dynamic state called the  {\it assimilated state}. The static state encompasses all model variables defined at start-time that are left unchanged by model propagation. The static state is the same for all  ensemble members. This may be the mesh topology used for spatial discretization, given it is invariant to model propagation and all members use the same.
The union of the static and dynamic state is the full set of information needed by the model to propagate an ensemble member further in time. The assimilated state is the sub-part of the dynamic state that is required by the update phase of the assimilation process  according to the available observations. We make this distinction as this is the base for the dynamic load balancing and  fault tolerance mechanisms of \melissa detailed in the following sections. Indeed, a \melissa runner only keeps the static state for its whole lifetime, while the  dynamic states of the members  are exchanged between the server and the runners.  The server can thus checkpoint a member saving its dynamic state only or assign a given member to a different runner by providing its dynamic state. Practically, properly identifying the dynamic state can be challenging for some applications.

\subsection{Server}

\begin{figure}[!ht]
    \centering
    \includegraphics[width=\textwidth]{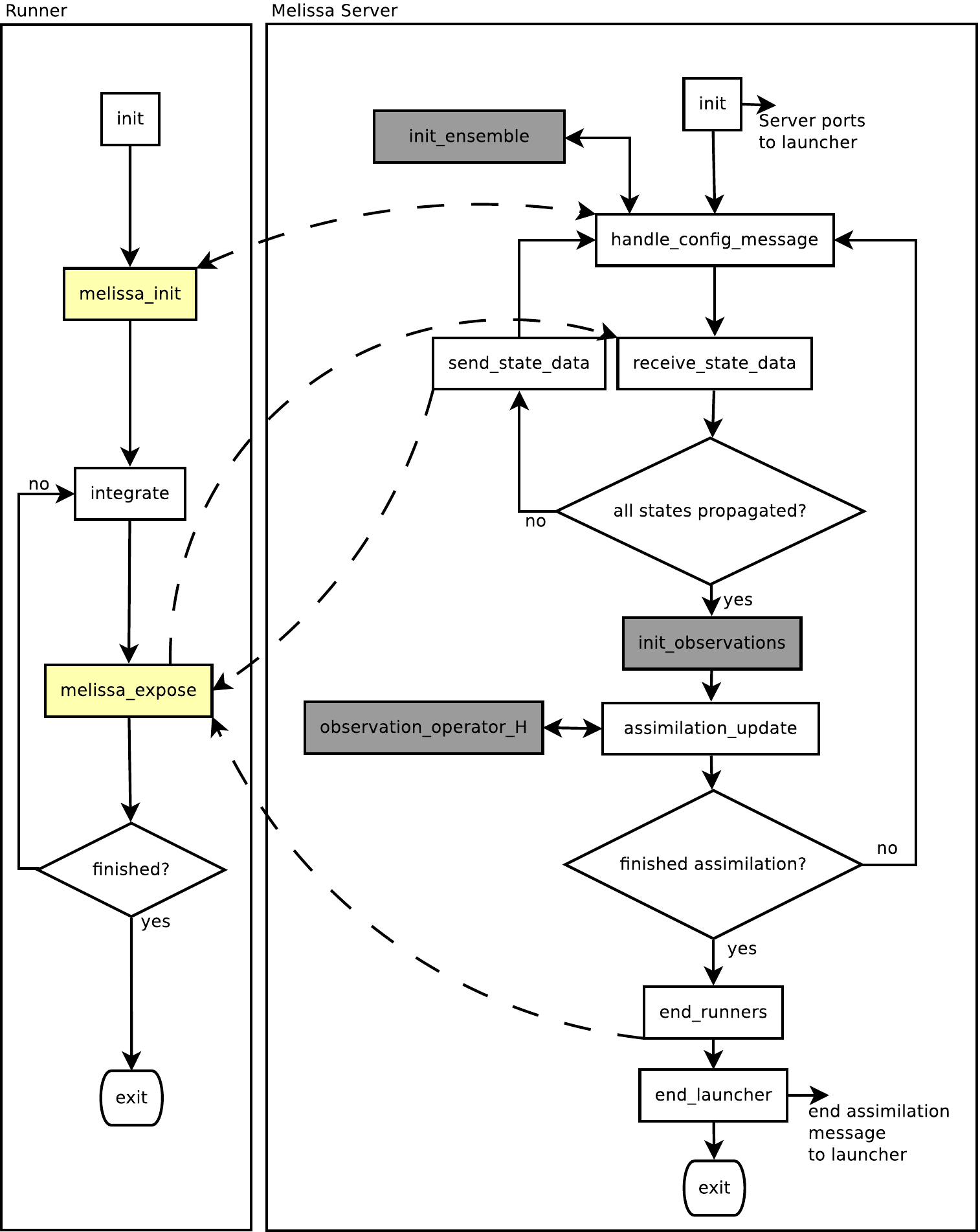}
    \caption{\melissa runner and server interactions (fault tolerance part omitted for sake of
      clarity). Dashed arrows denote messages that are exchanged between different components. Grey boxes are methods that need to be implemented by the user.
    Yellow boxes are \melissa API calls that need to be introduced in the simulation  code
    to transform it  into a runner.
   }
    \label{fig:impconcept}
\end{figure}

At every assimilation cycle the server collects the dynamic states of all ensemble members  from the runners to calculate the next ensemble of analysis states. At the same time it works as a master for the runners, distributing the next ensemble member (dynamic) states to be propagated to the runners. The server is parallel (based on MPI)  and runs on several nodes. The number of nodes required for the server is primarily  defined by its memory needs. The amount of memory needed is in  the order of the sum of  the member dynamic states. As detailed in \autoref{sec:asim_only_part}, the dynamic state (the state to be assimilated plus the extra simulation variables  that differentiate a member from the other) is the minimal amount of information needed to restore a given member on any runner.

The server needs to be linked against a user defined function to initialize all the member states (\autoref{fig:impconcept}  \texttt{init\_ensemble}). Other functions, e.g., to load the current assimilation cycle observation data (\texttt{init\_observations}) and the observation operator $\mathcal{H}$ (\texttt{observation\_operator\_H}) must be provided to the server for each DA study. In the current version
user functions are called sequentially by the server, but we expect to support concurrent calls to  \texttt{init\_ensemble} and \texttt{init\_observations} to further improve the server performance.

The current server embeds PDAF as parallel assimilation engine. The server parallelization  can be chosen independently from the runner parallelization. A  N~$\times$~M data redistribution takes place between each runner and the server to account for different levels of parallelism on the server and runner side. This redistribution scheme is implemented on top of  ZeroMQ, a connection library extending sockets~\cite{zmq_book}. This library supports a server/client connection scheme allowing dynamic addition or removal of runners.

Care must be taken to store  the coherent state vector parts and simulation variables together as they might not be received by all server ranks in the same order, as runners are not synchronized.  For instance server rank~0 could receive a part of ensemble state vector~3 while  rank~1 receives a part of ensemble state vector~4. Even more importantly the state parts that are sent back must be synchronized so
that the ranks of one runner receive the parts of the same ensemble state vector from all the
connected server ranks. For that purpose  all received state parts are labeled with the ensemble
member ID they belong to, enabling the server to assemble coherently distributed member states.
State propagation is ensured by the server rank~0, the only one making decisions on
which runner shall propagate which ensemble member. This decision is next shared amongst all the server
ranks using non blocking MPI broadcasts. This way communication between different server and runner ranks overlaps while other runner(ranks) perform unhindered model integration.

\subsection{Runners}\label{sec:runners}

\melissa runners are based on the simulation code, instrumented using the minimalist \melissa
API. This API consists only of two functions: \texttt{\meli\_init} and
\texttt{\meli\_expose} (\autoref{fig:impconcept}). \texttt{\meli\_init} must be called once at the beginning to define the size of the dynamic state vector per simulation rank.  This information is then exchanged with the server, retrieving the server parallelization level. Next  \texttt{\meli\_init}  opens all necessary connections to the different server ranks.

\texttt{\meli\_expose} needs to be inserted into the simulation code to enable extraction
of the runner  dynamic state and to communicate it with the server. When called, this function is given a pointer to the runner's dynamic state data in memory that is sent to the server who saves it as background state. Next \texttt{\meli\_expose} waits to receive from the server the dynamic part of an analysis state, used to update the runner  dynamic state. The function \texttt{\meli\_expose} returns the number of
timesteps the received analysis state shall be propagated or a stop signal.

\subsection{Launcher}

To start a \melissa application, a user starts the launcher that then takes care of setting up the server and runners on the supercomputer.

The launcher typically runs on the supercomputer  front node and is the only part of the \melissa
application that interacts with the machine batch scheduler. The launcher  requests resources for starting the server job and as soon as the server is up it submits  jobs  for runners. Hereby it prioritizes job submission within the same job allocation (if the launcher itself was started within such an allocation). Otherwise the launcher can also submit jobs as self contained allocations (by, e.g., calling \texttt{srun} outside of any allocation on Slurm based supercomputers). In the latter case it can happen that the server job and some runner jobs are not executed at the same time leading to inefficient small \melissa runs. Thus ideally at least jobs for the server and some runners are launched within the same allocation ensuring the \melissa application to operate efficiently even if no further runner jobs can be submitted.
It is further also possible to instruct the launcher to start jobs within different partitions. %

If the launcher detects that too few runners are up, it requests  new ones, or, once notified by the server that the assimilation finished, it deletes all pending jobs and finishes the full application.  The launcher also  periodically checks that the server is up, restarting it from the last checkpoint if necessary. The notification system between the server and the launcher is based on ZeroMQ. There are no direct connections between runners and the launcher. The launcher only observes the batch scheduler information on runner jobs.

\subsection{Fault tolerance}

\melissa supports detection and recovery from  failures (including straggler issues) of the runners through timeouts, heartbeats and server checkpoints. Since the server stores the dynamic states of the different members and each runner builds  the static state at start time,  no checkpointing  is required  on the runners. So \melissa ensures fault recovery even if the model simulation code  does not support checkpointing. If supported, runners can leverage  simulation checkpointing  to speed-up runner restart.

The server is checkpointed using the FTI library~\cite{bautista-gomez_fti_2011}, enabling  the recovery from server crashes without user interaction. The server sends  heartbeats  to the launcher.  If missing, the launcher  kills  the runner and server jobs and restarts automatically from the last server checkpoint.
The server is also in charge  of tracking runner activity based on timeouts. If  a runner is detected as
failing, the server re-assigns the current runner's member propagation to  an other active runner. More precisely, if one of the server ranks detects a timeout from a runner, it notifies the server rank~0 that reschedules this  ensemble member to a  different runner,  informing all  server ranks  to discard  information already received  by the failing runner.  Further, the server sends stop messages to all other ranks of the failing runner.  The launcher, also notified  of the failing runner, properly kills it and requests the batch scheduler to  start a new runner that will connect to the  server as soon as ready.

One difficulty are errors that cannot be solved  by a restart, typically numerical errors or, e.g., a wrongly configured server job.
To circumvent these cases \melissa counts the number of restarts. If the maximum, a user defined value, is reached, \melissa stops with an informative error message.  In the case of a recurrent error on a given member state propagation, it is possible to avoid stopping the full  application.  One option is to automatically replace such members with new ones by calling a user defined function for generating  new member states, possibly by perturbing existing ones.  Alternatively when the maximum number of restarts for a member state propagation is reached, this member could simply be canceled. As the number of members is high, removing a small number of members usually does not impair the quality of the DA process.  These solutions remain to be implemented in \melissa.

A common fault are  jobs being canceled by the batch scheduler once reaching the limit walltime. If this occurs at the server or runner level,
the fault tolerance protocol operates.

The launcher is the single point of failure. Upon failure  the application needs to be restarted by the user.

\subsection{Dynamic load balancing}\label{sec:dynamic-load-balance}

As already mentioned, runners send to the server the dynamic state of each member. For the  sole  purpose of DA, only  the sub-part of this dynamic state that we call the assimilated state  would  actually be necessary. But  having the full dynamic state on the server side brings an additional level of flexibility central  to the \melissa  architecture:  runners become agnostic of the members they propagate. We rely on this property for the dynamic load balancing mechanism of \melissa.

Dynamic load balancing is a very desirable feature when the propagation time of the different members differ.  This is typically the case with solvers relying on iterative methods, but also when runners are started on heterogeneous resources, for instance nodes with GPUs versus nodes without,
or if the network topology impacts unevenly the data transfer time between the server and runners.
The server has to wait  for the last member to return its background state before being able to proceed with the update phase computing the analysis states. The worst case occurs when state propagation is fully parallel, i.e.\ when each runner is in charge of a single member. In that case runner idle time is the sum of the differences between each propagation time and  the slowest one. As we target large numbers of members, each member potentially being a large-scale parallel
simulation, this can account for significant resource under utilization (\autoref{fig:traces} gives
a view of this effect for a simulated execution trace for ParFlow).
To reduce  this source of inefficiency \melissa enables 1)~to control the propagation concurrency level independently from the number of members 2)~to distribute dynamically members to runners.

The \melissa load balancing strategy relies on the \textsc{Graham} list~scheduling algorithm~\cite{graham_bounds_1966}.  The server distributes the members to runners on a first come first serve basis. Each time a runner becomes idle, the server provides it with the dynamic state of one member to propagate. This algorithm is simple to implement, has a very low operational cost, and does not require any information on the member propagation time.  The performance of the list~scheduling algorithm is guaranteed to  be at worst twice the one of the optimal scheduling that requires to know the member execution time.   More precisely  the walltime~$T_{ls}$  (called makespan in the scheduling jargon) is bounded by the optimal walltime $T_{opt}$:
\begin{equation}
    T_\text{ls} \le T_{opt}*(2-\frac{1}{m})
\end{equation}
where $m$ is the number  of  machines used, in our case the number of runners~\cite{shmoys_scheduling_1991}.
This  bound is tight, i.e.\ cannot  be lowered, as there exist instances where this bound is actually met.

A static scheduling distributing  evenly the members to runners at the beginning of each
propagation phase, does not guarantee the same efficiency as long as we have no knowledge on the member propagation time. The worst case occurs if one runner gets the members with the longest propagation time.

Also the list~scheduling algorithm is efficient independent of the number of runners, combining well with the \melissa runner management strategy. The number of expected runners is statically defined by the user at start time.  But the actual number of executed runners depends on the machine availability and batch scheduler. Runners can start at different time periods, they may not all run due to resource limitations, some may crash and try to restart. With list~scheduling,  a runner gets from the server the next member to propagate as soon as connected and ready.

From this base algorithm several optimizations can be considered. In particular data movements could be reduced by trying to avoid centralizing all dynamic states on the runner using decentralized extensions of list~scheduling like work stealing~\cite{Blumofe}. This could be beneficial when the assimilated state only represents a small fraction of the dynamic state. This is left as a future work.

\subsection{Code}\label{sec:code}

The code of \melissa (server and API) is written in C++, relying on features  introduced with
cpp14. This is especially handy regarding smart pointers to avoid memory leaks and having access to
different containers (sets, lists, maps) used  to store scheduling mappings. The assimilation
update phase  is contained in its  own class deriving the \texttt{Assimilator}-interface, which accesses the received dynamic states  and  creates a new set of analysis states to be propagated.

Implementing new ensemble-based DA methods within the \melissa  framework is straightforward, requiring to specify how to initialize the ensemble and how to transform the ensemble of background states  in an ensemble of analysis states using observations.

A derived class calling the PDAF  EnKF update phase  methods was implemented and is linked against the user defined methods to inititalize the ensemble and observations and to apply the observation operator (\autoref{fig:impconcept}).
From PDAF perspective, the \melissa server acts as a parallel simulation code assembling a "flexible assimilation system" with one model instance propagating all ensemble members sequentially online \footnote{see \url{http://pdaf.awi.de/trac/wiki/ModifyModelforEnsembleIntegration}, retrieved the 25.08.2020}. By handling the server  MPI communicator to PDAF, the update phase is parallelized on all server cores.

To let \melissa support other assimilation algorithms implemented in PDAF (e.g.,  LEnKF, LETKF\dots\footnote{for a complete list see
\url{http://pdaf.awi.de/trac/wiki/FeaturesofPdaf}, retrieved the 25.08.2020}) only  classes inheriting the \texttt{Assimilator} interface with calls to the desired PDAF filter update methods must be implemented.

The \melissa launcher is written in Python. To execute a \melissa study, a user typically executes a
Python script  for configuring the runs, importing  the launcher module and launching  the study.

The \melissa code base contains a test suite allowing end-to-end testing against results retrieved using PDAF as reference implementation. The test suite also  contains test cases validating recovery from induced runner and server faults.

The \melissa  code base will be made open source on GitHub\hide{\footnote{\sebastian{\url{https://github.com/razupaltuf/razupaltuf}TODO: put link!}}} upon publication. %

\section{Experimental study}\label{sec:exp}

Experiments in \autoref{sec:ultra_large_ens} and \autoref{sec:faultandelas} were performed on the Jean-Zay supercomputer on up to 424 of the  \numprint{1528} scalar compute nodes. Each node
has  \uu{192}{GB} of memory and two Intel Cascade Lake processors with 40 cores at \uu{2.5}{GHz}. The compute nodes are connected through an Omni-Path interconnection network with a bandwidth of \uu{100}{Gb/s}. The other experiments ran on the JUWELS supercomputer  (2 Intel Xeon processors, in total 48 cores at \uu{2.7}{GHz} and \uu{96}{GB} of memory per compute node, EDR-Infiniband (Connect-X4)) \cite{JUWELS}.

For all experiments we keep nearly the same problem size. Experiments assimilating ParFlow simulations use an assimilated state vector of \numprint{4031700} cells (pressure, \numprint{4031700} doubles, $\approx$~\uu{30.8}{MiB}), representing the Neckar catchment in Germany. For our test we were provided  observations from  25 ground water measuring sensors distributed over the whole catchment. Observation values were taken from a virtual reality simulation~\cite{schalge_presentation_2020}. The dynamic state contains the pressure,  saturation and density vectors for a total size of  $\approx$~\uu{92.4}{MiB}).  ParFlow runners are parallelized on one full node (40 processes for experiments on Jean-Zay and 48 processes for JUWELS respectively). For the experiments profiling the EnKF  update phase (\autoref{sec:enkfupd}), a toy model  from the PDAF examples, parallelized only on half of a node's cores, is used to save compute hours. For this experiment the dynamic state equals the assimilated state (\numprint{4032000} grid cells, $\approx$~\uu{30.8}{MiB}). The \texttt{init\_ensemble} function uses an online approach relying on one initial system state adding some uniform random noise for each member.
Loading terabytes of data for the initial ensemble states is thus avoided.
To further avoid the influence from file system jitter, assimilation output to disk is deactivated in the following performance measurements.

For the sake of simplicity and  minimal intrusion, the code was instrumented by calls to the STL-chrono library\footnote{see \url{https://en.cppreference.com/w/cpp/chrono}, retrieved the 24.06.2020}, allowing a precision of a nanosecond for these measurements. This minimal instrumentation was successfully validated against measurements using automatic Score-P instrumentation and Scalasca~\cite{knupfer_score-p_2012}.

\subsection{ParFlow}

For first tests we assimilate ParFlow simulations with \melissa. ParFlow is a physically based, fully coupled water transfer model for the critical zone that relies
on an iterative Krylov-Newton solver~\cite{ashby_parallel_1996,jones_newtonkrylov-multigrid_2001,maxwell_development_2005,kollet_capturing_2008,maxwell_terrain-following_2013}. This solver performs a changing amount of iterations until a defined convergence tolerance is reached at each timestep.

\subsection{Ensemble propagation}\label{sec:ens_prop}
\begin{figure}
    \centering
    \includegraphics[width=\textwidth]{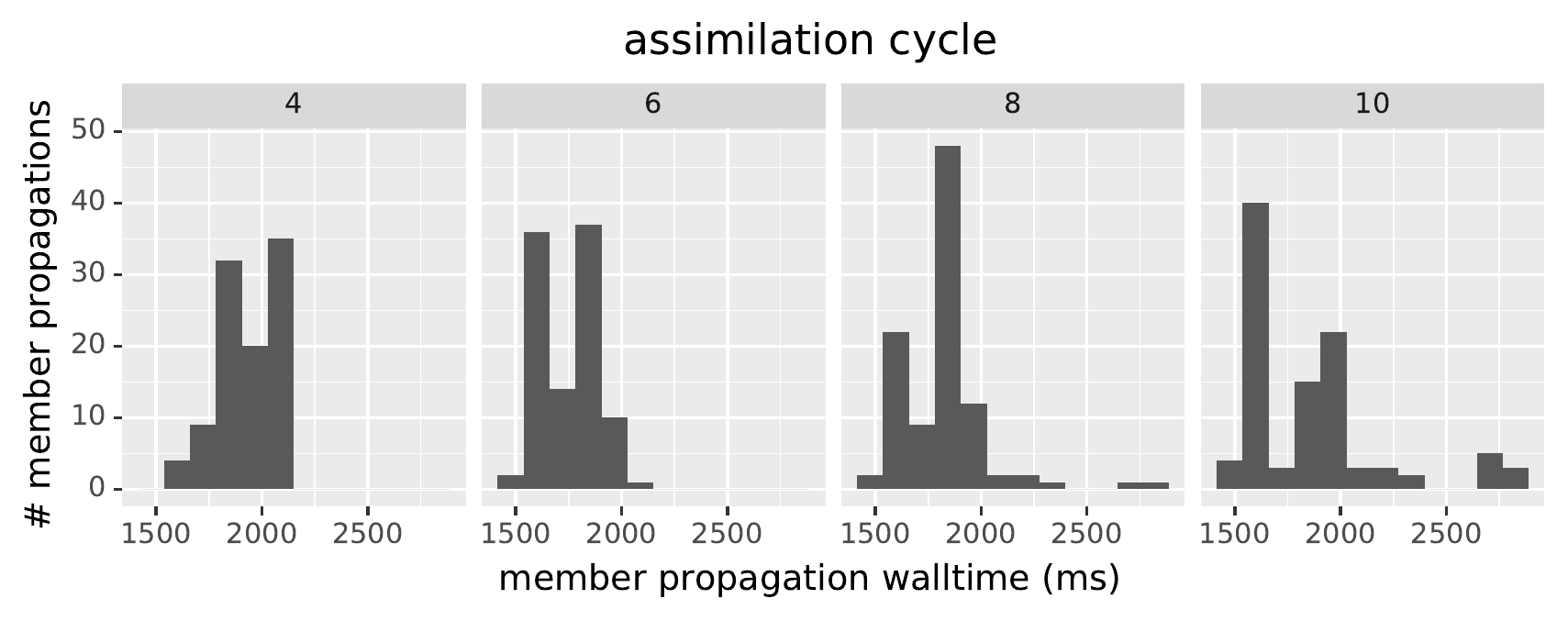}
    \caption{
        Histogram of propagation walltimes for 100 members during multiple assimilation cycles}
    \label{fig:propagate}
\end{figure}

\autoref{fig:propagate} shows the walltime distribution of 100 ParFlow member propagations for multiple  assimilation cycles.
The propagation time can vary significantly from about \uu{1.5}{s} to  \uu{2.5}{s}, with an average at \uu{1.9}{s}. The main cause for these fluctuations is  the Krylov-Newton solver used by ParFlow that converges  with different number of iterations depending on the  member state.
As detailed in \autoref{sec:dynamic-load-balance}, these variations  can impair the execution efficiency. \melissa mitigates
this effect by dynamically distributing members to runners following a list~scheduling algorithm.

\subsection{EnKF ensemble update}
\label{sec:enkfupd}

\begin{figure}
    \centering
    \includegraphics[width=\textwidth]{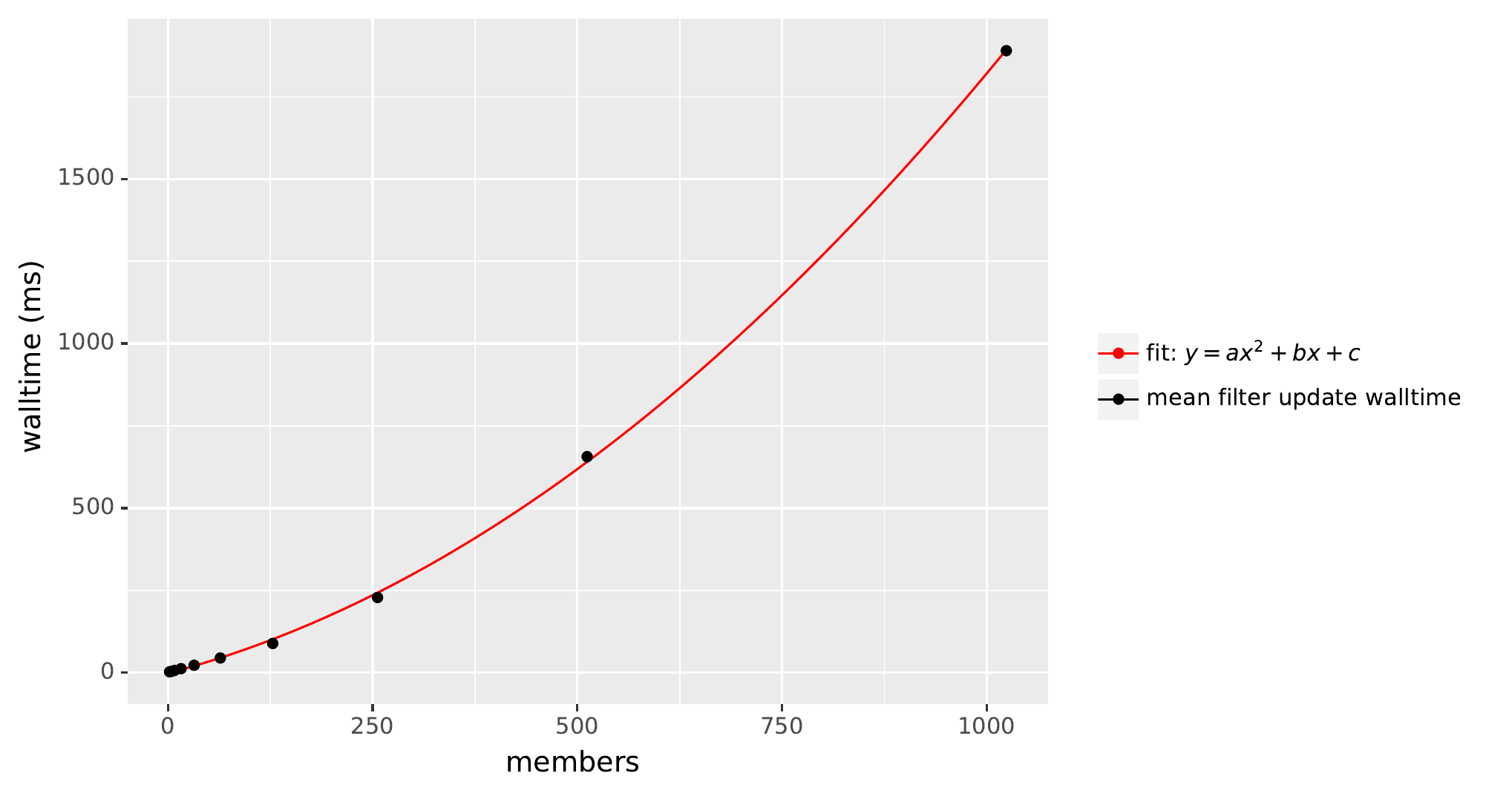}
    \caption{%
        Assimilating 288 observations into about \uu{4}{M} grid cells with up to \numprint{1024} members on JUWELS. Mean over 25 update phase walltimes.} %
    \label{fig:update}
\end{figure}

\autoref{fig:update} displays the evolution of the update phase walltime depending on the number of members,
 using \melissa with the PDAF implementation of EnKF. The mean is computed over 25 assimilation cycles, assimilating 288 observations each time.
 Standard deviation is omitted as not being significant ($<$~6\% of the update phase walltime). The EnKF update phase is executed on 3 JUWELS nodes (144 cores in total).
  The EnKF update phase relies on the calculation of covariance matrices over $M$ samples resulting in a computational complexity for the EnKF update phase of $O(M^2)$,  $M$ being the
amount of ensemble members. This is confirmed by  the experiments that fit a square function.
The walltime of the update phase also depends on the amount of observations.  In the following experiments  less observations are used, leading to an update phase of about only \numprint{1.1} seconds.  %

\begin{figure}
    \centering
    \includegraphics[width=\textwidth]{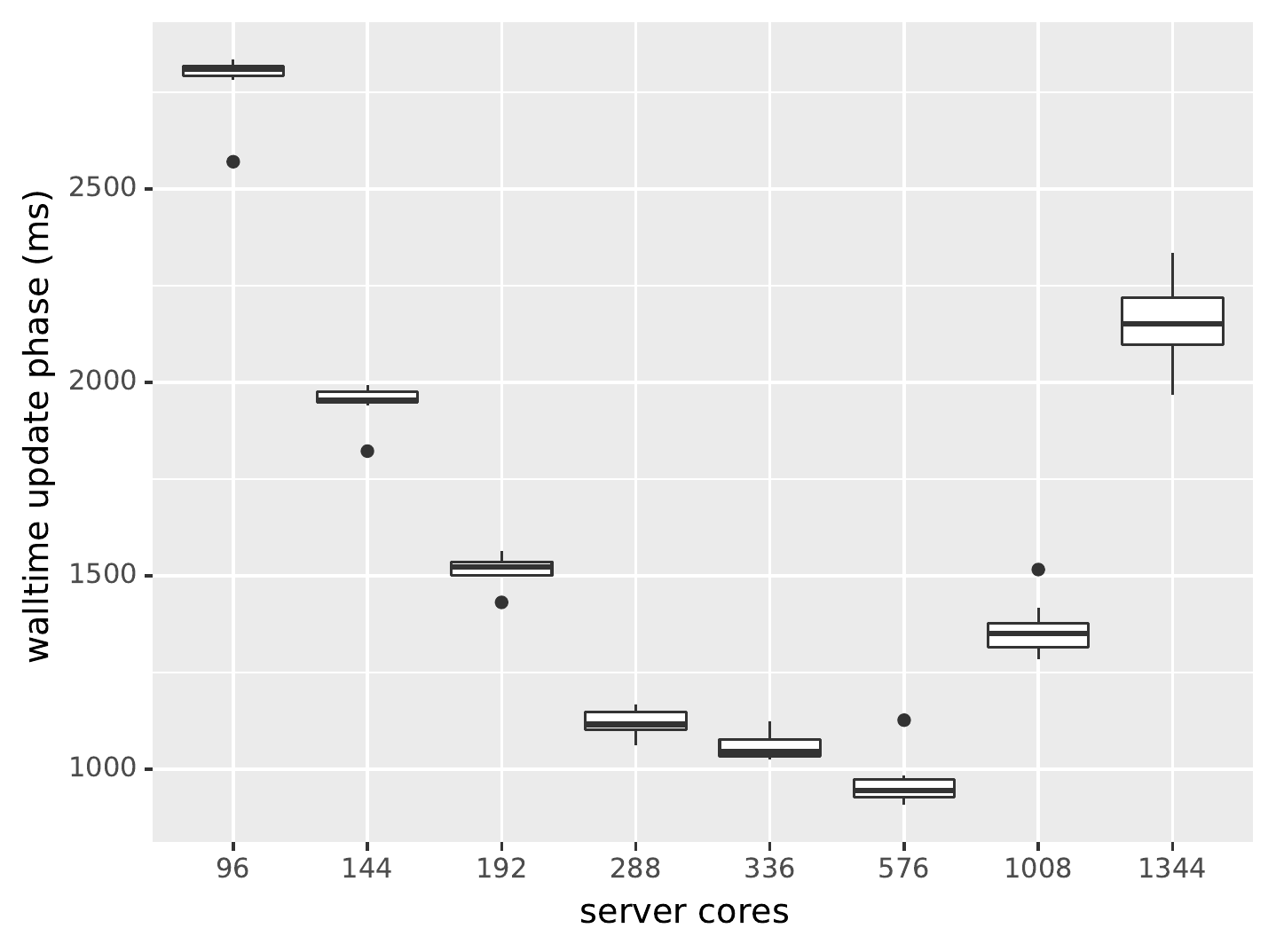}
    \includegraphics[width=\textwidth]{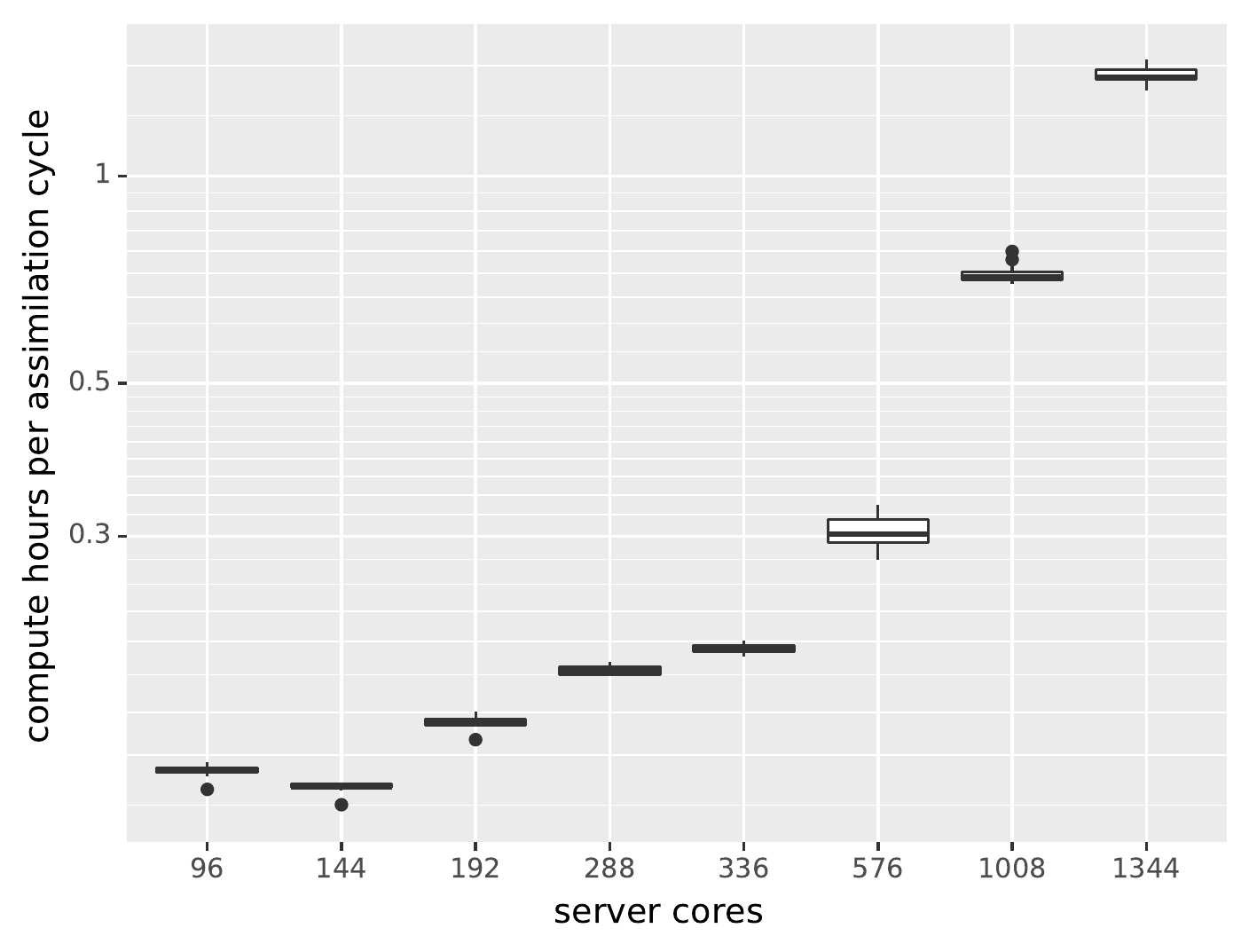}
    \caption{
        Boxplot of the update phase walltime (\textit{top}) and the total compute hours per assimilation cycle (compromising update and propagation phase, \textit{bottom}) when assimilating 288 observations into about \uu{4}{M} grid cells with \numprint{1024} members with a varying number of  server cores on  JUWELS.
        The latter graph can be used to evaluate server dimensioning.} %
    \label{fig:update-changing-cores}
\end{figure}

We also performed a strong scaling study (\autoref{fig:update-changing-cores}) timing the update phase for \numprint{1024} members and a varying number of server cores.
The parallelization  leads to walltime gains up to  $576$ cores.  Computing the covariance matrix for update phase is known to be difficult to efficiently parallelize.
Techniques like localization  enable to push the scalability limit. Localization is not used in this paper as we run with a limited number of observations.
As \melissa relies on PDAF which supports localization, localization can be easily activated by changing the API calls to PDAF in the \melissa \texttt{Assimilator} interface.  Refer to \cite{Nerger-PDAF-2013} and \cite{nerger_pdaf_2005} for further EnKF/PDAF scaling experiments.

Dimensioning the \melissa server optimally  depends on the assimilated problem dimensions, the used assimilation algorithm and the target machine. It should  be examined in a quick field study before moving to production. The results of such a field study can be seen in \autoref{fig:update-changing-cores}, \textit{bottom}. In the depicted case less core hours are consumed when not using a large number of server nodes (and cores respectively) since these idle during the whole propagation phase outplaying the walltime advantage they bring during the update phase. For the following experiments we assume that the update phase is short compared to the propagation phase leading to the  policy: Use the least server nodes possible to fulfill memory requirements. If the server would be faster using a sub-part of the cores available per node, use less cores (e.g., only 24 or 12 cores per server node). Some supercomputers provide special large memory nodes that could be leveraged to run the \melissa server.

\subsection{Runner scaling}\label{sec:runner_amount_dep_scaling}
\begin{figure}
    \centering
    \includegraphics[width=\textwidth]{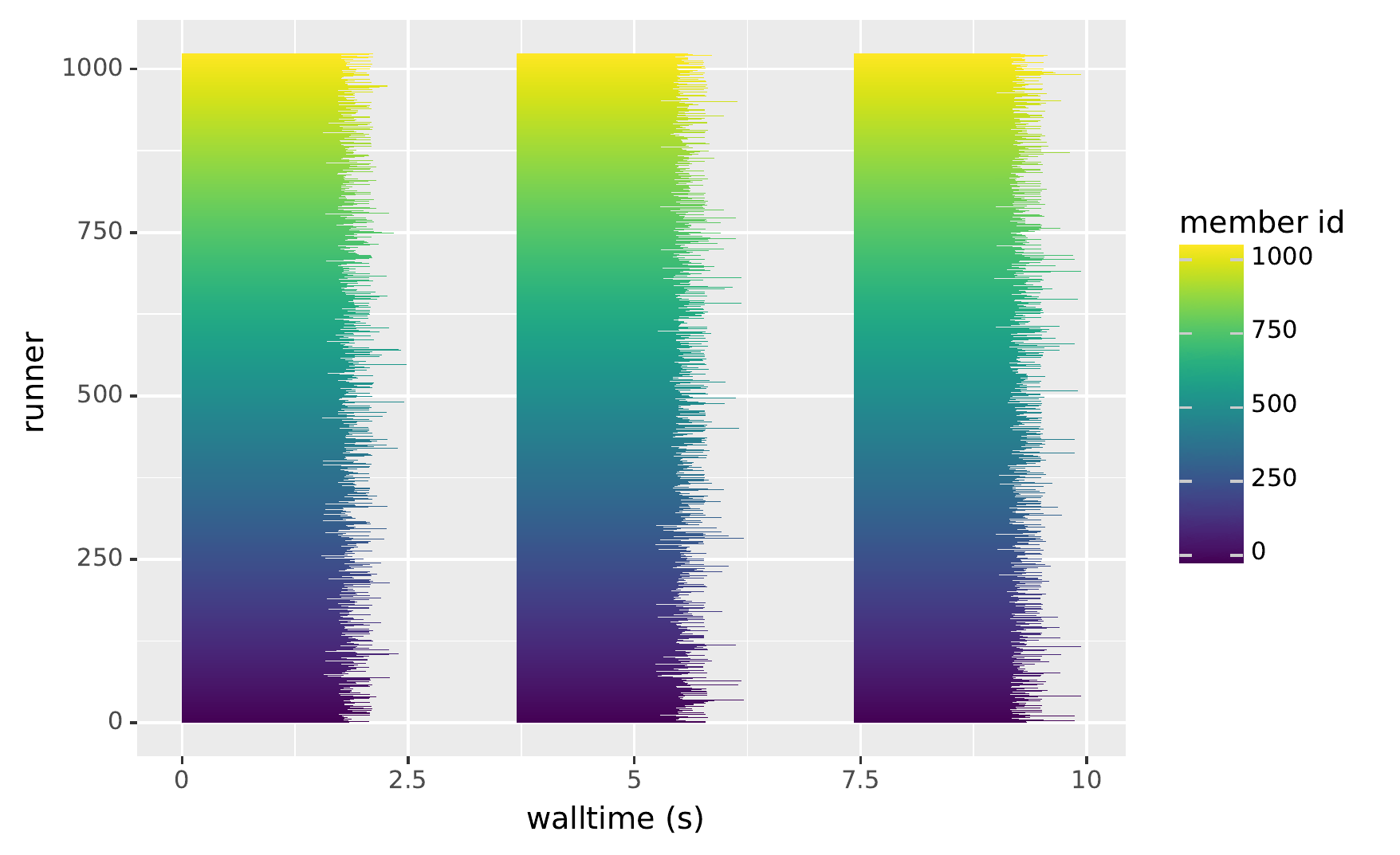}
    \includegraphics[width=\textwidth]{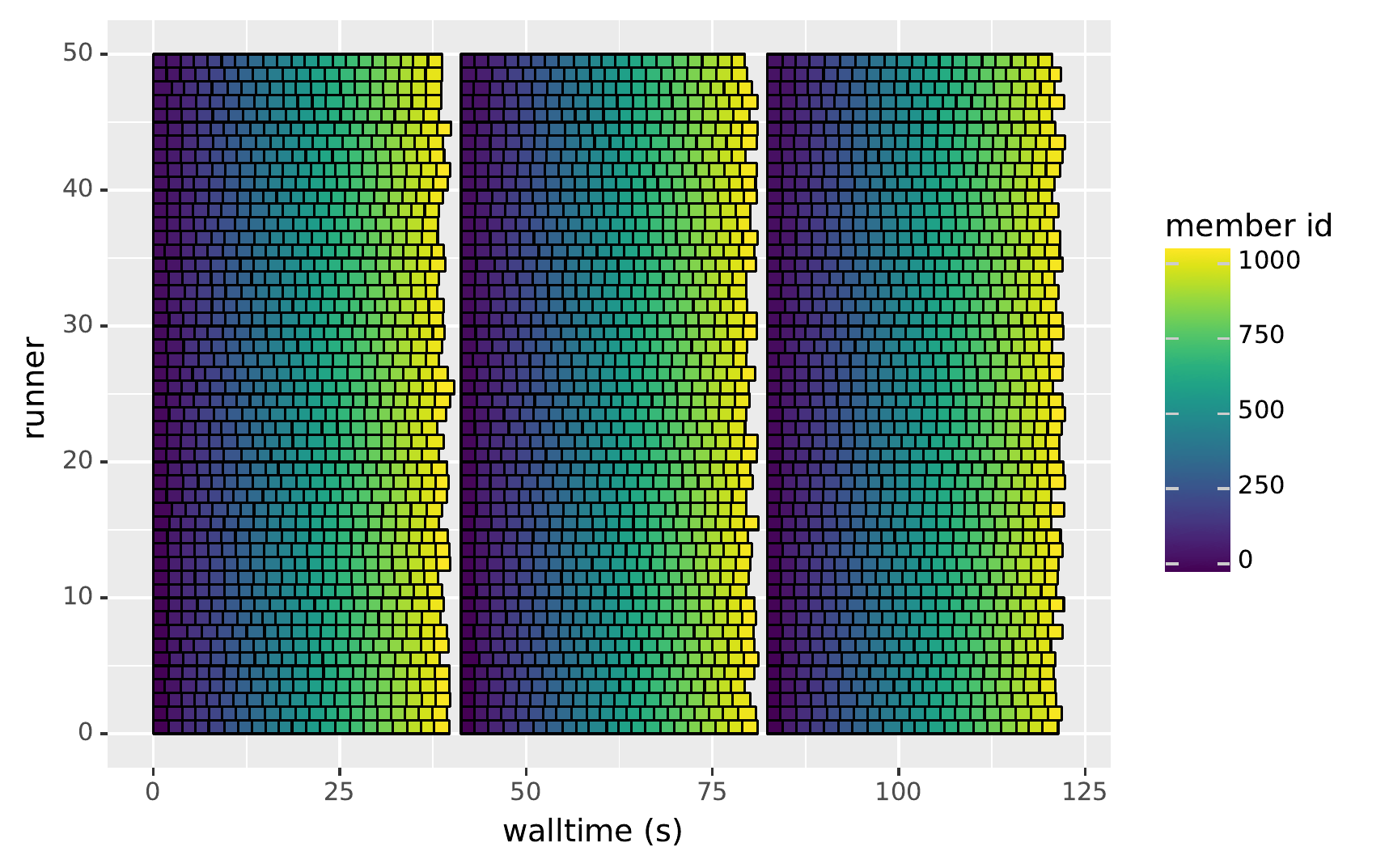}
    \caption{Load balancing \numprint{1024} members with list~scheduling on 50 (bottom)  and \numprint{1024} runners (top). Three assimilation cycles are plotted.  Simulated execution, based on member propagation walltimes from \autoref{fig:propagate}. Runners are idle about 50\% of the time when each one propagates a single member (top), while idle time is reduced to
    6\%  when each runner propagates about 20 members (bottom).
    }
    \label{fig:traces}
\end{figure}

\begin{figure}
    \centering
    \includegraphics[width=\textwidth]{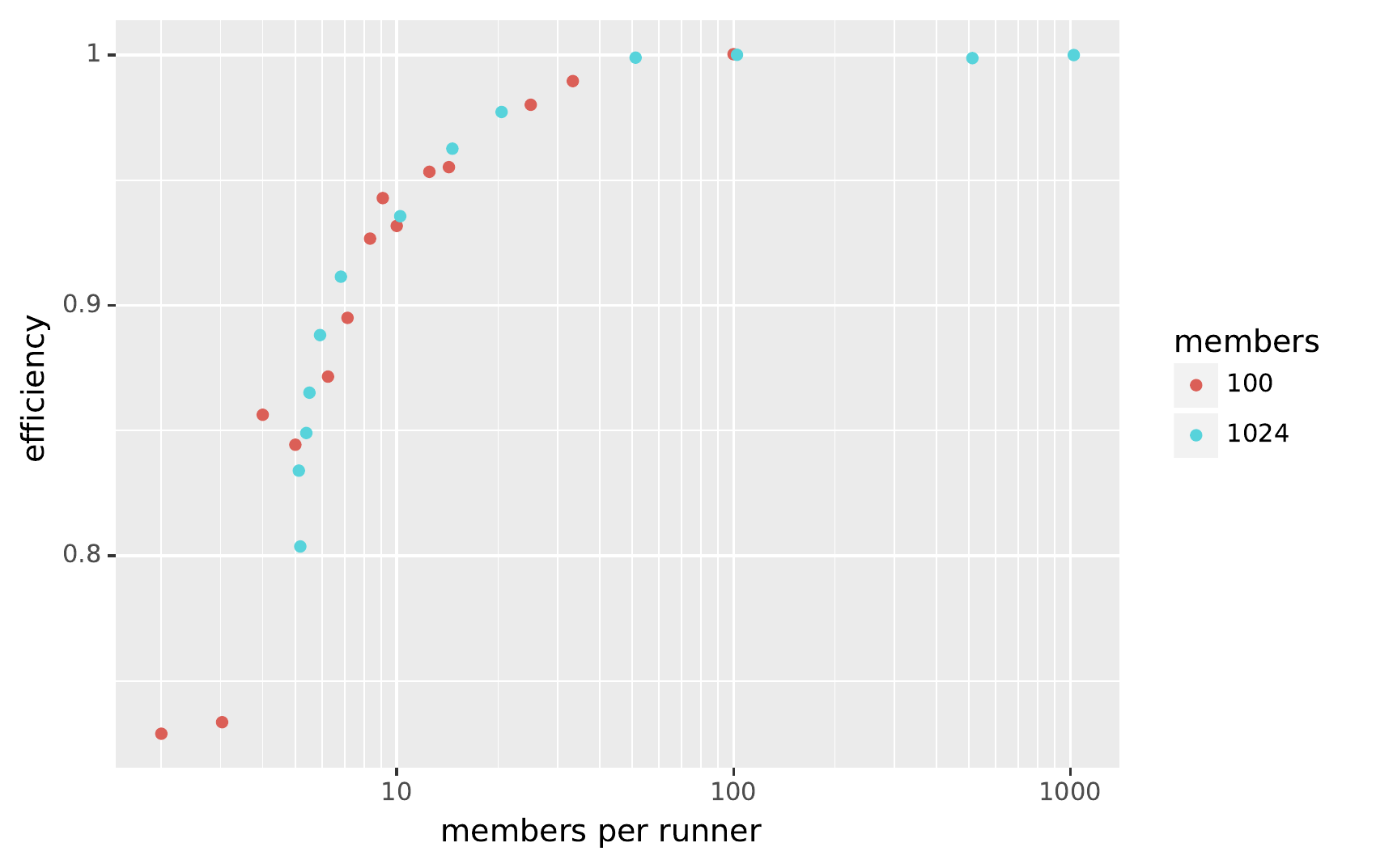}
    \includegraphics[width=\textwidth]{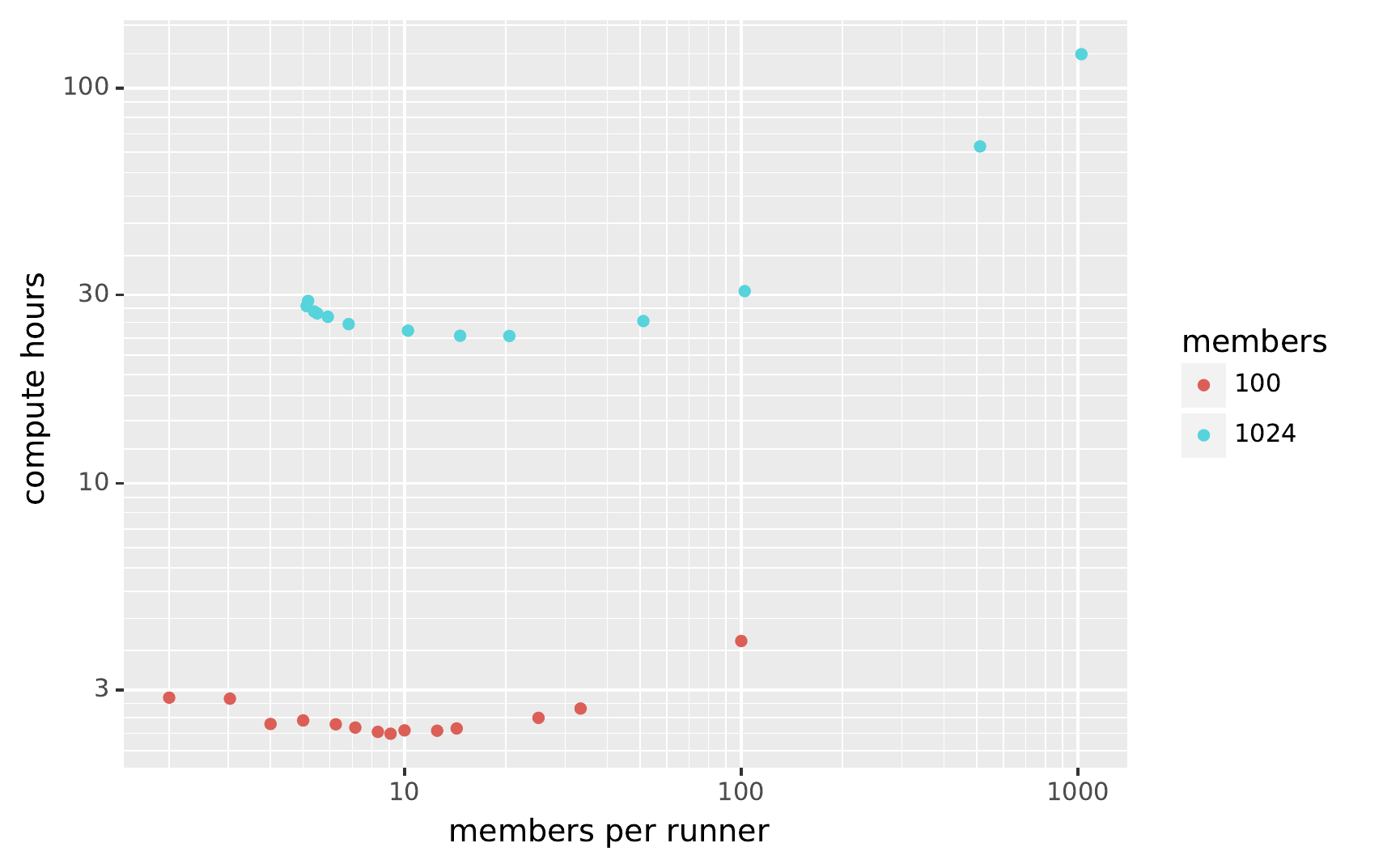}
    \caption{ Efficiency of the propagation phase only (\textit{top}) and total compute hours used per assimilation cycle (update and propagation phase) (\textit{bottom}) for different numbers of runners while assimilating 25 observations into an about \uu{4}{M} grid cell ParFlow simulation with 100 and \numprint{1024} ensemble members.}
    \label{fig:sweet_spot_1024}
\end{figure}

 We now focus on  the member to runner ratio.  A  single runner avoids idle runner time during the propagation phase, while having  as many runners as members ensures the shortest propagation time but  maximizes idle time  (\autoref{fig:traces}). Idle time also come from the switch between propagation and update phases: the server is mostly inactive during the propagation phase, while, in opposite runners are inactive during the update phase.

 We experiment with a varying number of runners for a fixed number of members (100 and \numprint{1024}) and a given server configuration (\autoref{fig:sweet_spot_1024}).  Plotted values result from an average  obtained from 8 executions, taking for each execution the time of the 2 last, over 3, assimilation cycles.  The efficiency of the update phase (\autoref{fig:sweet_spot_1024} top) is computed against the time obtained by running the members on a single runner.
 The compute hours are the total amount of consumed CPU resources (update and propagation phase, runners and server) during the assimilation cycles.  For both plots, standard deviations are omitted as
 being small (relative  standard deviations ($\frac{\text{standard deviation}}{\text{mean}}$) always smaller than $3\%$). The server was scaled to meet the memory needs. For the \numprint{100} (resp. 1024) member case, the server ran with 48 (resp. 240) cores on one (resp. 5) nodes. Each runner is  executed on 48 cores (1 node).

 Efficiency stays beyond $90\%$, when each runner propagates at least 7 or 8 members,
 $95\%$ for more than 10 members per runner and close to  $100\%$ for $50$ or $100$ members per runner. This demonstrates the efficiency of \melissa load balancing algorithm that maintains high efficiencies down to a relatively small number of members per runner. Obviously these
 levels of efficiency also depend on the distribution of propagation walltimes (see \autoref{sec:ens_prop}).

 But  the resources used for the server also need to be considered.  The total amount of compute hours (\autoref{fig:sweet_spot_1024}  bottom) shows a U shape curve with a large flatten bottom at about 9 members per runner for the 100 members case  and at about  20 for the \numprint{1024} members case: a sufficiently large  number of runners is required to amortize the server cost.  Changing the runner amount around those sweet spots changes significantly the efficiency of the update phase but slightly impacts the total  compute hours: the efficiency variation is  compensated by the impact on the server idle time during the update phase that varies inversely (the server is mostly idle during this phase). This also shows that  changing the number of runners in these areas is efficient. This advocates for leveraging \melissa elasticity for adding/removing runners according to machine availability.

 If we look at the compute hours, the $10\times$ increase in the number of members to propagate roughly matches the increase in compute hours. Thus the resource usage is here dominated by the propagation phase and the server does not appear as a bottleneck.

  The server exchanges about \uu{92}{MiB} of state data per member and assimilation cycle. Up to 24 (resp. 96) states are assimilated per second for the 100 (resp. \numprint{1024}) members case (propagation \textit{and} update phase) using about 2 (resp. 5) members per runner.

\subsection{Ultra-large ensembles} \label{sec:ultra_large_ens}

\begin{figure}
    \centering
    \includegraphics[width=\textwidth]{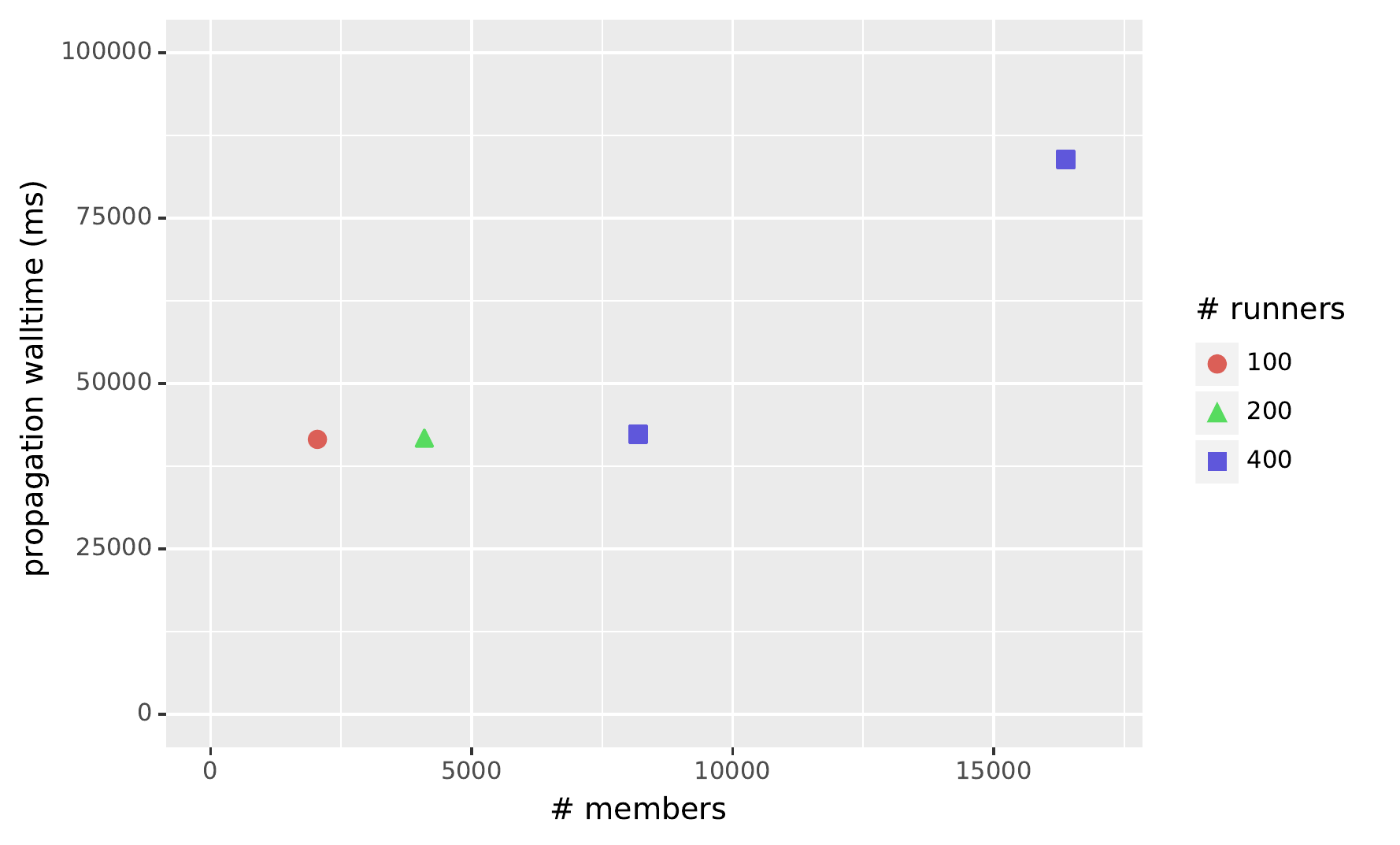}
    \caption{Scaling the assimilation of the ParFlow domain on up to \numprint{16384} members}
    \label{fig:scaling-many-members}
\end{figure}

In this section we scale \melissa to ultra-large ensembles, assimilating with up to \numprint{16384} members. To save compute hours only  a few assimilation cycles  ran on up to \numprint{424} compute nodes using up to \numprint{16240} cores of the Jean Zay supercomputer. When doubling both, the ensemble size and the number of runners, the execution time stays roughly the same, with an average number of about \numprint{20} members per runner. The number of runners was not increased when running \numprint{16384}  members, as we were not able to get the necessary resource allocation on the machine. Thus, the walltime  doubles.   Notice that  we plot only the second assimilation cycle as the first cycle  shows higher execution times as not all runners are   connected to the server yet.

For each assimilation cycle with \numprint{16384} members, a total of \uu{2.9}{TiB} of dynamic state data (\uu{0.96}{TiB} being assimilated data) are transferred back and forth over the network  between the  server and all the runners. By enabling direct data transfers, \melissa avoids the performance penalty that would induce the use of files as intermediate storage.

The server is spread on the minimum number of nodes necessary to fulfill the memory requirements (\uu{1.9}{TiB}
to hold the dynamic states of the \numprint{16384} ensemble members). Since \melissa and the underlying PDAF assimilation engine parallelization are based on domain decomposition and the domain size does not change when the number of members goes up, the number of allocated server processes is kept constant at \numprint{240} cores. Thus the update phase takes a considerable amount of the application walltime (up to \uu{53}{s}  for the \numprint{16384}  members case).  %
Future work will look at using techniques like localization to reduce the cost of the update phase. As being supported by PDAF, this should not entail any modification to \melissa.

\subsection{Fault tolerance and elasticity}\label{sec:faultandelas}

\begin{figure}
    \centering
    \includegraphics[width=\textwidth]{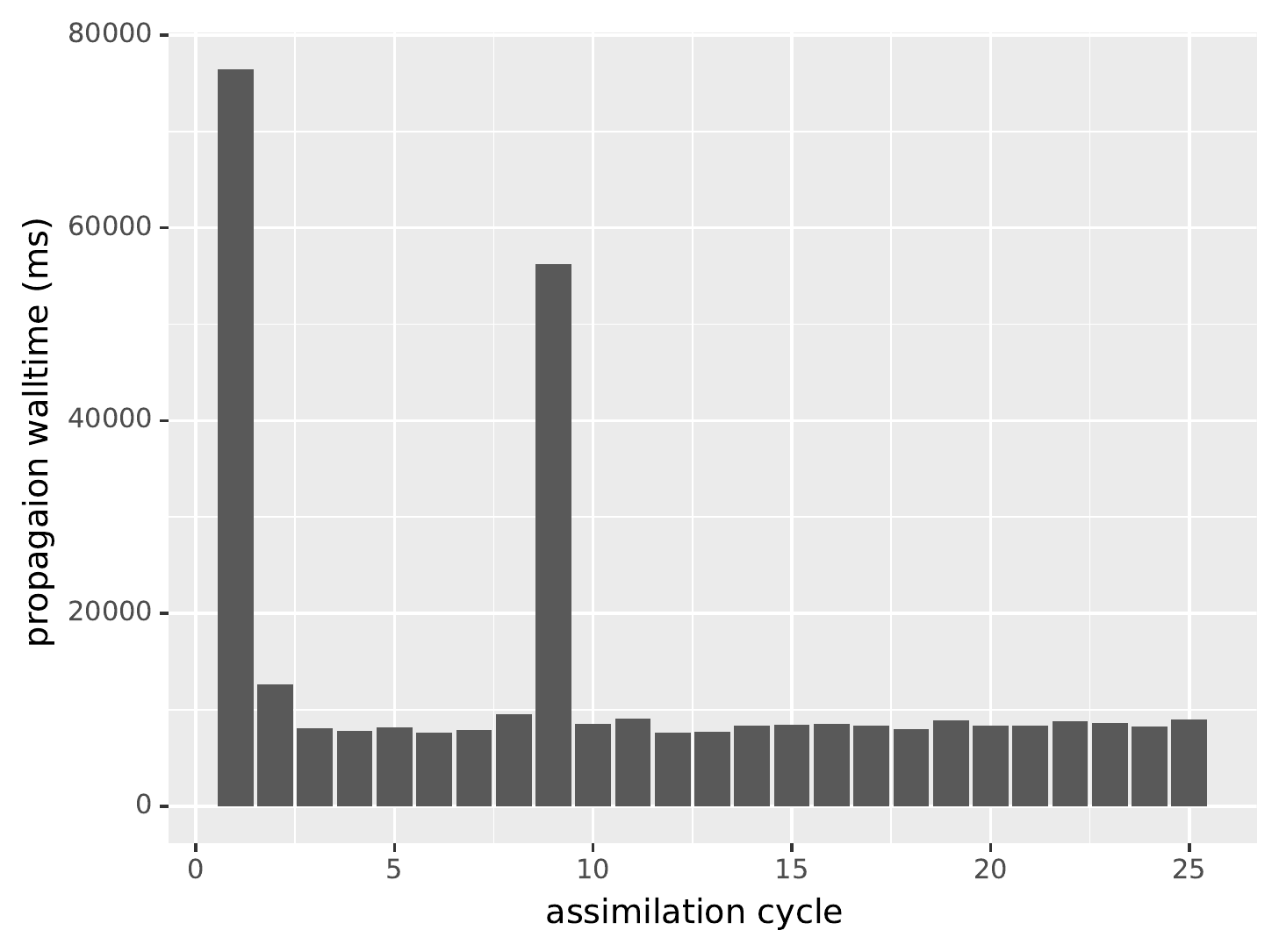}
    \caption{Walltime of the propagation phase over different cycles for one ParFlow assimilation (\numprint{30} ensemble members, \numprint{9} runners at the beginning, 1 server node). Higher propagation walltimes are due to a lower number of active runners not all ready from the beginning (first and second cycles) or after a crash ($9^\text{th}$ cycle).}
    \label{fig:crashtest}
\end{figure}
Runners can be dynamically added or removed, giving the \melissa elasticity.
Also the fault tolerance protocol benefits of this feature. This simple experiment on one execution aims to demonstrate the \melissa elasticity (\autoref{fig:crashtest}). The first and second propagation phase take longer since not all runners are ready to accept members from the beginning (respectively about \uu{80}{s} and \uu{13}{s}) instead of less than \uu{10}{s}.  At the  $9^\text{th}$ assimilation cycle, 4 out of 9 runners are killed. The timeout for runner fault detection was set to \uu{25}{s}. When the server detects the runner faults, their member propagations are rescheduled to other runners. At the same time the launcher also detects these faults and restarts 4 runners.  Restarting took place within \uu{55}{s} after the runner crashes. Already at the $10^\text{th}$ propagation phase 9 runners are propagating members again.

Notice that here we leverage the batch scheduler capabilities (Slurm). A single allocation encompassing all necessary resources  is requested at the beginning. Next jobs for the server or runners are allocated by Slurm within this envelope, ensuring a fast allocation. When the runners crash, again the restarted runners reuse the same envelope. A hybrid scheme is also possible, requesting a minimal first allocation
to ensure the data assimilation to progresses fast enough, while additional runners are allocated outside the envelope but whose availability to accept members may take longer depending on the machine load.
It is planned to rely in such situations on \textit{best effort} jobs that are supported by batch schedulers like OAR~\cite{oarpaper}. Best effort jobs can be deleted by the batch scheduler whenever resources to start other higher prioritized jobs are needed. This way \melissa runners can fill up underutilized resources between larger job allocations on the cluster.
All these variations on the allocation scheme require only minimal customization of the assimilation study configuration.

\section{Conclusion}\label{sec:conclusion}

In this article we introduced \melissa, an elastic, fault-tolerant, load balancing, online framework for ensemble-based DA. All these properties lead to an architecture allowing to run Ensemble Kalman Filters with up to  \numprint{16384}~members to assimilate observations into large-scale simulation state vectors of more than \uu{4}{M}~degrees of freedom. At the same time \melissa scaled  up to \numprint{16240} compute cores.

Thanks to the master/worker architecture adding and removing resources to the \melissa application is possible at runtime. In future we want to profit from this agile adaptation to further self-optimize the  compute hour consumption.
We will also consider executions on heterogeneous machines (nodes with accelerators or more memory). The modular master/worker model of \melissa allows for flexibility to leverage such architectures. Large memory nodes for the server can also have a very positive impact on the core hour consumption. We are  also planning  to integrate other assimilation methods into \melissa and to run use cases with diverse more complex simulation codes and at different scales.

Concerned with measuring the compute cost and environmental impact  of this paper, we counted the total number of CPU hours used, including all intermediate and failed tests that do not directly appear in this paper, at about \numprint{136000} CPU hours split between the JUWELS and the Jean-Zay supercomputers.

\section*{Acknowledgements}
  We would like to thank the following people:  Bibi Naz, Ching Pui Hung and  Harrie-Jan Hendricks Franssen from Fohrschungszentrum Juelich for the scientific exchange as well as for providing the ParFlow real world use case for data assimilation,
  Kai Keller and Leonardo Bautista-Gomez from Barcelona Supercomputing Center for the integration of FTI into the \melissa server code as well as
  Lars Nerger from Alfred Wegener Institut and Wolfgang Kurtz from Leibniz Supercomputing Center for the scientific exchange on PDAF,
  ParFlow and TerrSysMP. We also thank the DataMove research engineers Christoph Conrads and Théophile Terraz
 for contributing to the \melissa software stack and proof reading.

\section*{Funding}
 This project has received funding from the European Union's Horizon 2020 research and innovation program under grant agreement No 824158 (EoCoE-2).  This work was granted access to the HPC resources of IDRIS  under the allocation 2020-A8  A0080610366 attributed by GENCI (Grand Equipement National de Calcul Intensif).

\bibliographystyle{plain}
\bibliography{biblio}

\end{document}